# Asset Allocation Using a Markov Process of Clustered Efficient Frontier Coefficients States

Nolan Alexander[1*], William Scherer[1], and Jamey Thompson[2]

[1] Systems Engineering, University of Virginia, Charlottesville, VA
[2] Managing Member/Portfolio Manager at Jocassee Quantitative, New York, NY
[*] Corresponding Author, email: nolan_alex2018@yahoo.com

**Abstract**

We propose a novel asset allocation model using a Markov process of states defined by clustered efficient frontier coefficients. While most research in Markov models of the market characterize regimes using return and volatility, we instead propose characterizing these states using efficient frontiers, which provide more information on the interactions of underlying assets that comprise the market. Efficient frontiers can be decomposed to their functional form, a square-root second-order polynomial defined by three coefficients, to provide a dimensionality reduction of the return vector and covariance matrix. Each month, the proposed model hierarchically clusters the monthly coefficients data up to the current month, to characterize the market states, then defines a Markov process on the sequence of states. To incorporate these states into portfolio optimization, for each state, we calculate the tangency portfolio using only return data in that state. We then take the expectation of these weights for each state, weighted by the probability of transitioning from the current state to each state. To empirically validate our proposed model, we employ three sets of assets that span the market, and show that our proposed model significantly outperforms benchmark portfolios.

**1. Introduction**

The information that efficient frontier coefficients provide has its foundation in Modern Portfolio Theory (MPT). MPT, developed by Harry Markowitz (1952), uses mean-variance quadratic programming to optimally allocate portfolio weights that minimize volatility. The model provides a pareto frontier consisting of all risk-return optimal portfolios, which is known as an efficient frontier (EF). The theoretical best portfolio on the EF is the portfolio that maximizes the Sharpe ratio, which is the ratio of excess return to standard deviation (Sharpe, 1966). This Sharpe-maximizing portfolio is known as the tangency portfolio (Tobin, 1958).

The MPT model makes multiple assumptions, and a number of researchers have developed extensions to relax these assumptions. These include adding transaction costs (Davis and Norman, 1990), and extending to continuous time (Merton, 1975). The primary assumption that this paper will focus on is that the market is stationary. We can easily observe that the market is nonstationary, as future expected asset returns and covariances can be significantly different from historical data. This assumption often leads to significant estimation errors in practice, which limits the robustness of the Markowitz model. There already exist multiple models that attempt to relax this assumption, with some of the most notable being the seminal works of the Black-Litterman model, and covariance matrix estimation models.

Black and Litterman (1992) proposed a Bayesian model to provide superior posterior estimates of the return vector and covariance matrix. The Black-Litterman model combines two priors: views implied by the market, and investor views that are provided as input.

To better estimate the covariance matrix when the number of assets is larger than the number of samples, which causes instability in estimation, Ledoit and Wolf (2003) proposed a covariance shrinkage model. Engle and Sheppard (2001) proposed modeling the autocorrelation of the covariance terms to provide superior forecasts of the covariance matrix.

While these models can improve estimation of MPT parameters, they still have limitations. The Black-Litterman model can be difficult to use as it requires users to input their own market estimates and uncertainty as matrices and vectors, which must be determined external to the model itself. The improved covariance matrix estimation models cannot provide superior estimates of the return vector, which is the parameter that the mean-variance model is most sensitive to.

Separate from research in portfolio optimization, other researchers have developed Markov regime-shifting models. Most regime-shifting models are Hidden Markov Model (HMM), which are models that assume that each point in time is dependent on an unobserved state that follows a Markov process. An HMM fits a model that maps points in time to states by maximizing an objective function such as the loglikelihood function. HMMs have been applied to a number of financial datasets and features including economic data (Hamilton, 1989), stock index returns (Hardy, 2001), individual stock returns (Hassan and Nath, 2005), return and the distance from the covariance matrix to the identity matrix (Burkett et al., 2019), and the Mahalanobis distance calculated using the precision matrix and return vector (Procacci and Aste, 2019).

A number of researchers have developed extensions to the HMM for financial data. Dai et al. (2010) develop a trading strategy using an HMM with two states as a forecaster of trend. Hsu et al. (2012) developed an HMM that uses spectral clustering to determine states. Bae et al. (2014) developed a portfolio optimization model under an HMM framework. Aydınhan et al. (2023) added a penalty to the HMM objective function to reduce jumps and generalize the hidden state variable to a vector of probabilities of being in each regime.

While there have been many developments in regime-shifting, these models still often have poor performance out-of-sample compared to forecasting models that do not use regime-shifting, as HMMs are sensitive to small misclassifications (Dacco and Satchell 1999).

While researchers less commonly apply unsupervised models on financial data, a more popular area of research is utilizing supervised learning models to forecast the market and develop asset allocation strategies (Nevasalmi, 2020; Pinelis and Ruppert, 2022; Daul et al., 2022). These models assume that there exists only one regime.

We apply our regime-shifting model to a dataset that has not been applied to in literature thus far: EF coefficients. The use of EF coefficients in modelling is related to previous works. The standard approach to creating an EF is to use mean-variance optimization to find several efficient portfolios and extrapolate between these, but Merton (1972) derived a closed-form solution of the EF as a square root second-order polynomial function with three coefficients. Alexander and Scherer (2023) defined a set of interpretable EF coefficients that can be used to forecast tangency portfolios and developed an asset allocation model using these tangency portfolio forecasts. Alexander and Scherer (2023) also used the interpretable coefficients in a supervised machine learning approach to forecast the market direction. While both this paper and the previous work of Alexander and Scherer used the EF coefficients as a feature set, there are key differences in the approaches. Both the previous works focused on using the EF in a *supervised* approach. The first work forecasted EFs and used those forecasts to develop improved portfolios through geometric properties of the tangency portfolio, while the second work forecasted the direction of the entire S&P 500 with a decision tree model. However, this current paper uses the EF coefficients as features in an *unsupervised* approach to develop clustered market states, and uses these states and state transitions to develop improved portfolios under a Markov process.

Unlike most regime-switching models, our proposed Markov model uses observable states rather than hidden states, which allows our model to be more interpretable. Our proposed Markov model defines states with a clustering model similar to Hsu et al. (2012), but for clustering, we only use cluster distance in the objective function, rather than fitting to a Markov model. Our proposed Markov model also applies portfolio optimization under regime-switching similar to Bae et al. (2014), but we use an observable Markov model rather than an HMM.

We provide a novel approach to relax the MPT assumption of stationarity through using a Markov process of market states defined by clusters of EF coefficients. Rather than forecasting the MPT parameters, we propose that the stationary assumption holds better within the regimes determined by the clustering. At each time-step in forecasting the next state, we first hierarchically cluster the coefficients over all monthly intervals up to the current month to define the states, then define a Markov process on the sequence of states. Then, to construct a portfolio using the information provided from this Markov process, we develop a mapping of each state to the tangency portfolio weights created using only data in that state. We then take the expectation across all of the state's weights by the probability of transitioning from the current state to each state.

This paper provides two primary contributions. (1) This paper characterizes market states that follow a Markov process using a feature set that has not yet been applied to regime-shifting, the interpretable EF coefficients: $r_{MVP}$, $\sigma_{MVP}$, and $u$. (2) This paper also proposes a more interpretable alternative Markov asset allocation model to HMM portfolio optimization. The proposed model creates an optimal portfolio using the Markov process of observable clustered states and a mapping of states to tangency portfolios defined by data only in those states. We provide empirical results demonstrating that this novel approach significantly outperforms all benchmarks over time for three asset universes (Figures 7, 8, and 9). To help explain our proposed model in further detail, we will first explain two core concepts: EF coefficients, and temporal clustering.

## 1.1. Efficient Frontier Coefficients

Merton (1972) derived a functional representation of the EF as a square root second-order polynomial with three coefficients using Lagrange multipliers:

$$A = e^T V^{-1} e > 0 \quad (1)$$
$$B = r^T V^{-1} e$$
$$C = r^T V^{-1} r > 0$$

The equation for the EF is

$$\sigma^2(r) = \frac{Ar^2 - 2Br + C}{AC - B^2} \qquad (2)$$

Alexander and Scherer (2023) previously developed three interpretable coefficients:

$$r_{MVP} = \frac{B}{A}, \sigma_{MVP} = \frac{1}{\sqrt{A}}, u = \sqrt{\frac{AC - B^2}{A}} \qquad (4)$$

The equation for the EF with these interpretable coefficients is

$$\sigma^2(r) = (u^{-1}(r - r_{MVP}))^2 + \sigma_{MVP}^2 \qquad (3)$$

$r_{MVP}$ and $\sigma_{MVP}$ are respectively the return and standard deviation of the minimum variance point (MVP). $u$ is the rate of curvature of the EF utility function. An EF with a higher $u$ diminishes more slowly, and therefore has better tradeoffs of risk and return at every efficient portfolio except for the minimum variance point. These interpretable coefficients each control one graphical component of the EF: each coordinate of the vertex, and the rate of curvature. An increase in $r_{MVP}$ indicates greater returns for all levels of risk, which would suggest that the market is demanding more risk premium for all levels of risk. An increase in $\sigma_{MVP}$ indicates greater risk for all returns, which would imply a more volatile market. An increase in $u$ would signal a better market for risk-seeking investors.

### 1.2. Temporal Clustering

Some of the most common clustering models are k-means, agglomerative hierarchical clustering, and spectral clustering (Jain, 1999). K-means clusters using Euclidean spheres from cluster centroids. Agglomerative hierarchical clustering joins nearby points into classes that can become subclasses of higher-level classes until the highest-level class includes all points. Spectral clustering performs clustering on the eigenvectors of the Laplacian matrix of the data, which treats each point as a vertex on a graph. K-means only performs well when the clusters can be separated by spheres, and Spectral clustering can model much more complex separations between clusters. However, Spectral clustering is more sensitive to noise than hierarchical clustering, which is still able to model more complex cluster separations than k-means. As financial states likely follow a taxonomy, we selected hierarchical clustering for our proposed model's clustering method.

Mantegna (1999) shows that hierarchical clustering can be used to cluster assets returns based on the correlation matrix. As correlation is not a valid distance metric because it does not obey triangle inequality, we can convert correlation to a valid distance metric with the following transformation:

$$d_{|\rho_{ij}|} = \sqrt{2(1 - \rho_{ij})} \qquad (5)$$

These clustering algorithms are generally used for non-temporal data. To account for the time-series aspect of the data, we utilized Dynamic Time Warping (DTW). DTW measures the similarity of multiple time-series, and is commonly used in speech recognition to account for varying speaking speeds (Berndt and Clifford, 1994). While using Euclidean distance to measure similarity between two time-series is restrictive as data can only be compared when they occur at the same time, DTW is more flexible. For two time series $x$ and $y$, at time $i$, Euclidean distance can only compare $x_i$ and $y_i$, whereas DTW can compare $x_i$ and $y_j$ where $i \leq j$. DTW performs a monotonically increasing mapping of each time-series index to other time-series indices that minimizes the total cost, which is the sum of a distance function between the values of each mapped pair of indices.

If we calculated the DTW distance matrix on data with rows for each time interval and columns for each state-variable, it would output a symmetric matrix with rows and columns of each state-variable that shows the similarities between state-variable time-series, but not the similarity between time intervals, which is what we need. To calculate a DTW distance matrix with rows and columns for each time interval, we can calculate DTW on the correlation distance matrix of the time-series state-variables.

## 2. The Proposed Markov Markowitz Model

Our proposed model has five main steps. (1) Determine market states through the temporal clustering of monthly EF coefficients. (2) Define a Markov model from the state transitions. (3) For each state, calculate tangency portfolios using only data from that state. (4) Compute expected weights by weighting each set of state weights by the transition probability. (5) Repeat in an iterative, online updating fashion.

### 2.1. Defining Market States with Clustering

To define market states with EF coefficients, we elected to use hierarchical clustering. We selected hierarchical clustering as market regimes do not necessarily cluster in spheres, and Spectral clustering is less interpretable. Also, the hierarchical clustering algorithm aligns with the structure of the data: the market is either bullish or bearish, and there are different types of bullish and bearish markets.

Because we have time-series data, we performed DTW hierarchical clustering. To perform DTW hierarchical clustering of states we performed the following five steps: (1) Transpose the data such that the rows are each EF coefficient and columns are each month. (2) Compute the correlation matrix, such that each row and column of the matrix are each month. (3) Transform the correlation matrix to be a valid distance metric using equation 5. (4) Calculate the DTW distance matrix from the correlation matrix. (5) Perform hierarchical clustering using the DTW distance matrix.

### 2.2. Markov Model of Market State Transitions

The online Markov model iteratively re-clusters states each month using the EF coefficients data up to the current data. Then, the model defines a transition matrix from that clustering of states. We use the following notation:

$\tilde{t}_i \in [\tilde{t}_0, \dots, T]$ = current test time period at iteration $i$ in the expanding online framework
$X_{t,a}$, $t \in [0, \dots, \tilde{t}_i - 1]$ = train EF coefficient data indexed at time $t$ and asset $a$
$\tilde{X}_{t,a}$, $t \in [\tilde{t}_i, \dots, T]$ = test EF coefficient data indexed at time $t$ and asset $a$
$C_K$ = clustering model with $K$ clusters
$s_t \in [1, \dots, K]$ = states indexed at time $t$ defined when trained on data up to time $\tilde{t}_i - 1$
$P_{i,j}$ = transition matrix with the probability of transitioning from state $i$ to $j$ defined when trained on data up to time $\tilde{t}_i - 1$

The clustering model constructs a many-to-one mapping of months to states.
$$C_K(X) = s$$
We then assume the states follow a Markov process.
$$P_{i,j} = \mathbb{P}[s_t = i | s_{t-1} = j], \forall i \in [1, \dots, K], j \in [1, \dots, K]$$
Because the model is online, after each iteration, the current test time period moves forward with $\tilde{t}_{i+1} = \tilde{t}_i + 1$. Then, both $s$ and $P$ are recomputed with the new data from time $\tilde{t}_i$ now added to the train set.

## 2.3. The Tangency Portfolio Within Each State

To integrate the information from the Markov structure to an asset allocation model, we develop a bijective mapping of states to portfolio weights. At each time step, we determine the market states using clustering, and then partition the data by state. For each state partition, we calculate the tangency portfolio using only the return data in that state.

Unlike in standard MPT, where the tangency portfolio is calculated using *all* of the data regardless of state, the proposed model instead calculates a separate set of weights for each state using data from *only* that state. To handle regime shifts, or state transitions, MPT generally sets a specific lookback period or exponential decay factor to help mitigate the chances of including no-longer-relevant data. Because the proposed model defines market states that it partitions by, we can use the entire data with a full lookback, because each state weight uses only data in that state, so all of the data is relevant.

To select the state tangency portfolio, we run both optimizations for when the weights sum to 1 (long-only) and 0 (long-short), and select the portfolio among the two with the greater in-sample Sharpe using only data from that state. Let $\bar{r}_s$ and $V_s$ denote the return vector and covariance matrix respectively calculated using only the return data within state $s$. Formally, we have

$$(W_{s,:})_t := \arg\max_{w_s} \frac{w_s^T \bar{r}_s - r_f}{(w_s^T V_s w_s)^{1/2}} \quad \forall s \in [1, \dots, K]$$

where $(W)_t$ is a matrix with rows for each state and columns for each asset calculated at time $t$.

Now, we have a separate set of portfolio weights for each of the $K$ states. We then aggregate these $K$ portfolio weights into a single set of optimal weights. We do not know which state the market will go to in the next month, but the transition matrix provides the probabilities of entering each state based on the current state. We define the optimal weights as the expectation of the weights in each state, reweighting by the probability of transitioning to that state. Let $\boldsymbol{P}_{s_t,:}$ denote the transition matrix subset at the state for the current time, which is the discrete probability distribution vector of transferring to each state given the current state. We define the optimal set of weights as

$$\boldsymbol{w}_t^* \coloneqq \boldsymbol{P}_{s_t,:}^T (\boldsymbol{W})_t \tag{6}$$

We refer to this optimal set of weights as the Markov Markowitz weights.

**2.4. Online Updating**

To update the model with new information, the model is updated in an expanding online fashion, so that each month, the model uses all samples up to the current point in time. We first train the clustering algorithm and calculate mapping of states to weights on the current training data. Then, each month, we iteratively update the training data with the current data, and rerun the clustering and recalculate the Markov Markowitz weights.

We selected this updating method because in practice, after running the model in the current month, when running the model one month later, we will acquire new information for another month that we can incorporate into the model. Online updating also helps ensure that the backtest performance of the proposed model is not sensitive to the selection of the initial test start date.

**2.5. Example of the Proposed Markov Markowitz Model**

To help illustrate the proposed model, we provide a simplified example in figure 1. In this example, there are only two states, five months of data, and four assets. Given that we are in state 2 for May 2000, there is a 1/3 probability of transitioning to state 1 and a 2/3 probability of transitioning to state 2 in June 2000 based on the historical transitions of the Markov process. Therefore, the Markov Markowitz weights to hold through June 2000 are calculated as $\boldsymbol{w}^* = \frac{1}{3}\boldsymbol{W}_{1,:} + \frac{2}{3}\boldsymbol{W}_{2,:}$.

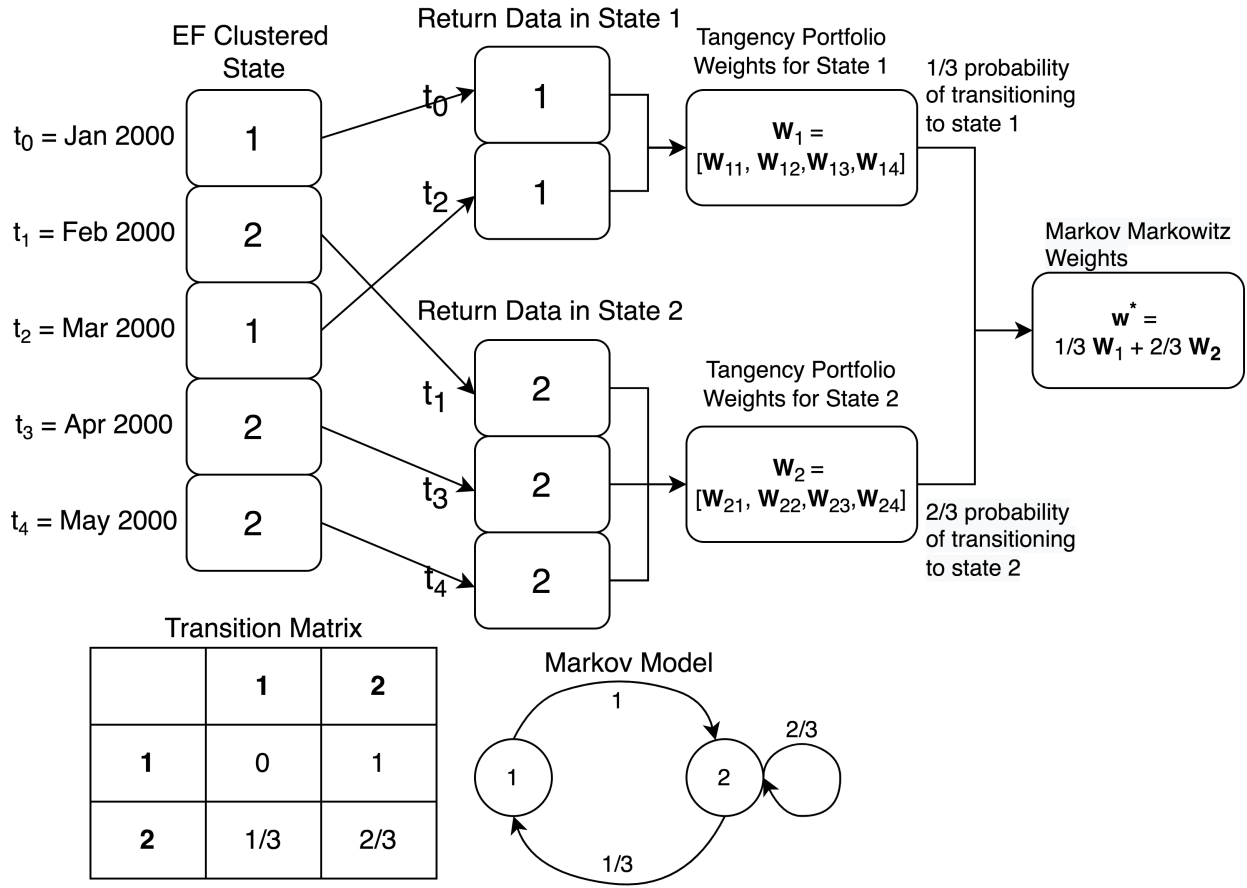

**Figure 1.** A visualization of a simple example of the proposed Markov Markowitz Model. In this example, there are only two states, five months of data, and four assets. Given that the last state is state 2, the probability of transitioning to state 1 is 1/3 and the probability of transitioning to state 2 is 2/3. The tangency portfolios for each state is calculated using only data within that state, and the final portfolio is an aggregation of these state tangency portfolio, weighted by the state transition probabilities: 1/3 and 2/3.

To summarize the entire process, figure 2 provides a visualization of the model process, with our novel contributions in green.

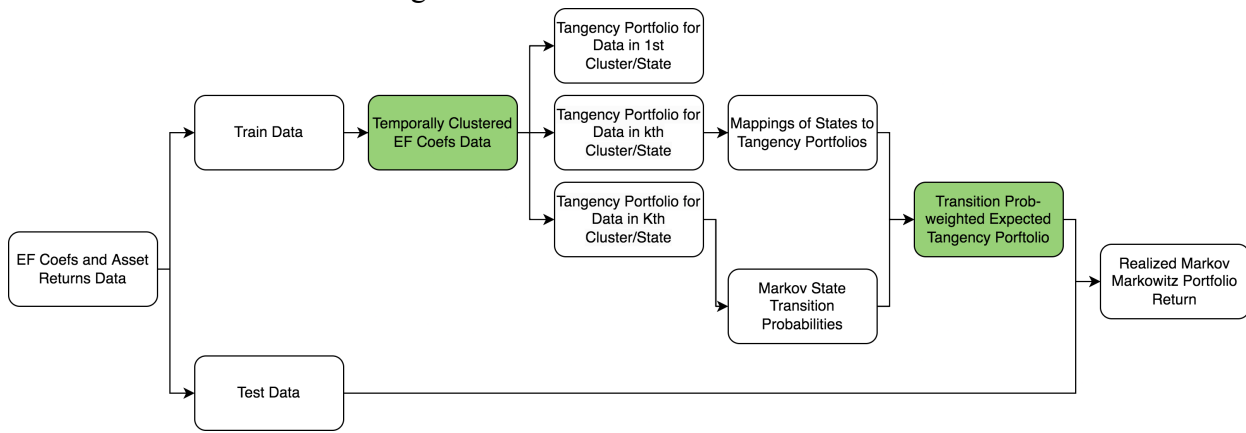

**Figure 2.** A visualization of the entire process of our proposed Markov Markowitz model, with novel contributions in green. The first is the use of EF coefficients to define market states with clustering. The second is the calculation of tangency portfolios within each cluster, and the aggregation by state transition probability weighting.

## 3. Data

### 3.1. Asset Universes

To test these dynamic Markov state portfolio models, we used three universes of assets. The first universe corresponds to mutual funds categorized by the Fama French 3-factor model's classification (Fama and French, 1993) of growth/value and market capitalization (we will refer to this universe as GVMC). The second universe corresponding to the 9 original S&P market sectors. The third universe consists of 5 global developed market indices. Table 1 shows the assets in each universe.

| Universe 1: Growth, Value, and Market Cap (GVMC) | | |
|---|---|---|
| **Asset Name** | **Description** | **Ticker** |
| Fidelity OTC Portfolio | large-cap growth | FOCPX |
| Fidelity Growth & Income Portfolio | large-cap value | FGRIX |
| Fidelity Growth Company Fund | mid-cap growth | FDGRX |
| Fidelity Low-Priced Stock Fund | mid-cap value | FLPSX |
| Invesco Oppenheimer Discovery Fund | small-cap growth | OPOCX |
| Heartland Value Fund Investor class | small cap value | HRTVX |
| **Universe 2: S&P Sectors** | | |
| **Asset Name** | **Description** | **Ticker** |
| Materials | Raw materials (e.g. iron, lumber) | XLB |
| Energy | Fossil fuels and renewables (e.g. oil, solar) | XLE |
| Finance | Banks, investment firms, insurance | XLF |
| Industrial | Tools for producing other goods (e.g. manufacturing machinery) | XLI |
| Technology | Software and hardware | XLK |
| Consumer Staples | Essentials (e.g. foods and beverages) | XLP |
| Utilities | Basic amenities (e.g. electricity, water) | XLU |
| Health Care | Medical services, equipment and insurance | XLV |
| Consumer Discretionary | Non-essentials (e.g. luxury goods, leisure activities) | XLY |
| **Universe 3: Developed Markets** | | |
| **Asset Name** | **Description** | **Ticker** |
| S&P 500 | United States | SPX |

| Hang Seng | China | HSI |
| --- | --- | --- |
| DAX | Germany | GDAXI |
| CAC 40 | France | FCHI |
| TSX | Canada | GSPTSE |

**Table 1.** The data within each of the three asset universes.

For the developed markets universe, we could not find data for the developed market indices with total return, so both the proposed model and benchmarks will appear to have significantly lower performance than if the futures were traded as in practice. These asset datasets were collected using the Yahoo Finance API. In addition, for the risk-free rate of the tangency portfolios, we used the French-Fama 3-factor data from Dr. French's website (2023).

The feature set that we used, the EF coefficients, are only defined for a lower frequency than the return data. For our analysis, we used the monthly EF coefficients calculated with daily return data. The monthly EF coefficients are calculated by taking the expected returns and covariance matrix of the daily returns within each month of daily returns. For example, to calculate the EF coefficients for January 2000, we compute the expected returns and covariance of the 21 days of return data in January 2000 for all the assets in the universe.

We selected a monthly time interval for the data because quarterly and yearly data did not provide enough data points. We could not select a more granular time interval, such as daily, because calculating the EF coefficients requires the computation of a covariance matrix. We cannot compute a daily covariance matrix without intraday data. We also cannot compute a rolling covariance matrix for these coefficients, because it would make the features that define states contain information from outside its time interval, making the states bleed together. For example, if we calculate the EF coefficients for December 2000 with a three-month rolling window, the EF coefficients would contain information from October and November 2000. This would cause the EF coefficient to not purely define the state for the month of December 2000, but also partially the prior two months.

### 3.2. Initial Train Test Splits

Although the backtests are performed in an online expanding fashion, we must determine an initial test split based on the length of the initial window. We cannot start the expanding window with only the first year of data, as that would only provide 12 data points, which would not provide enough data to define a Markov model of clustered states. As the backtests are expanding, these test start dates will move forward over time. For example, the GVMC universe test set starts on January 2000, and after computing the weights for January 2000, the model will add the January 2000 data to the train set, and the test set will now begin on February 2000. Because the backtest is online, the performance with a split at a further point in time will have the same returns after the split to if the model was run starting at that split, less the accumulated profit up to that point. For example, an online model with the test split starting in 2008 will have the same returns as an online model with the test split starting in 2000 for all performance after 2008, which is where the two test sets overlap. The only difference between the performance of the two online models with different test start dates is the level, as the model that started in 2000 would have accumulated profit from 2000-2007, so it would start at a higher level than the model that started in 2008, but both would increase by the same returns starting from 2008 on, with a log scale. Because of this property of the online backtesting procedure, the selection of the test start dates does not have a significant impact on the model performance as long as the initial training set is long enough for the model to learn initially.

To determine the initial test start date, we considered the tradeoff between not including enough data for the model to initially train on and not including enough data to measure performance out-of-sample. For each asset universe, the model must be trained on enough data to have seen both bull and bear markets, and must include a market crash so the model can include extreme events in its state model. At the same time, we would like to leave as many extreme events out-of-sample so we can see how the model performs in difficult environments. While it is possible that the proposed model could perform well during crashes by chance, intentionally excluding a crash from the test set by starting the test set after the crash would not allow us to measure the robustness of the proposed model. In addition, we can later observe in the backtest time-series plots that the outperformance of the proposed model does not all come from good performance during a single crash.

As the three datasets start at different times, we split on different dates for each universe. Both the GVMC and the Developed Markets universes go back over 10 years further than the Sectors universe. We decided to split the GVMC and Developed Markets universes before the Dot-com bubble, in the year 2000. If we also split the Sectors universe on 2000, it would only have one year of data, which would not be enough for training. Instead, we decided to split the Sectors universe before the 2008 recession. To align with the Sectors universe, we also performed a second split for each of the GVMC and Developed Markets universe at the same 2008 split to help ensure that the split selection did not bias the results.

|  | **Train Start** | **Test Start** | **Test End** |
|---|---|---|---|
| GVMC | 1990 | 2000, 2008 | 2022 |
| Sectors | 1999 | 2008 | 2022 |
| Dev Markets | 1988 | 2000, 2008 | 2022 |

Table 2. Train and test years for each universe. The GVMC and Developed Markets universes have two test start dates as we ran the backtest for both splits.

### 3.3. Hyperparameter Selection: Number of Clusters

The only hyperpameter of the proposed model is the number of clusters. To determine the number of clusters, we plotted a dendrogram showing the hierarchy when clustered with only the initial train data. Figures 3-5 show the dendrogram for each of the three universes.

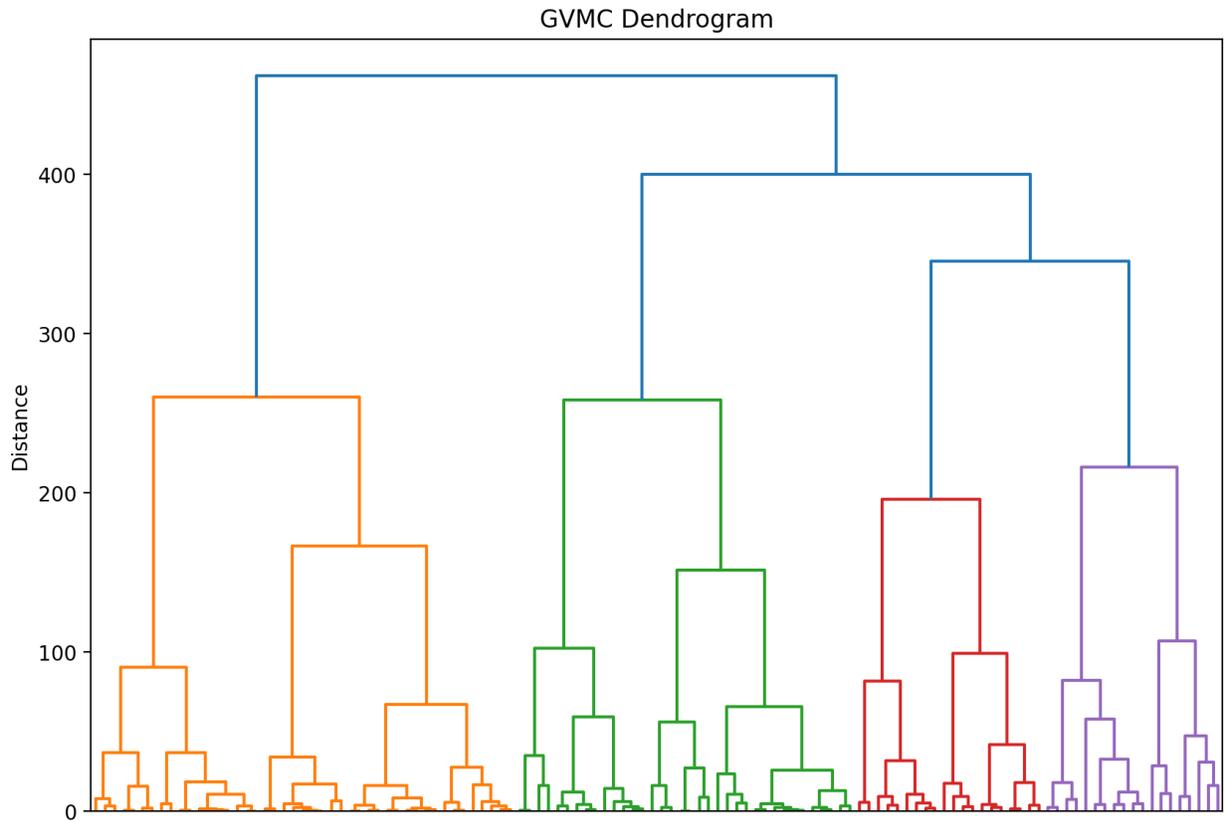

**Figure 3.** Dendrogram created from the DTW hierarchical clustering of the EF coefficients with the GVMC universe on the initial training data up from 1990 to 1999. The dendrogram shows that there are four primary clusters for the GVMC universe.

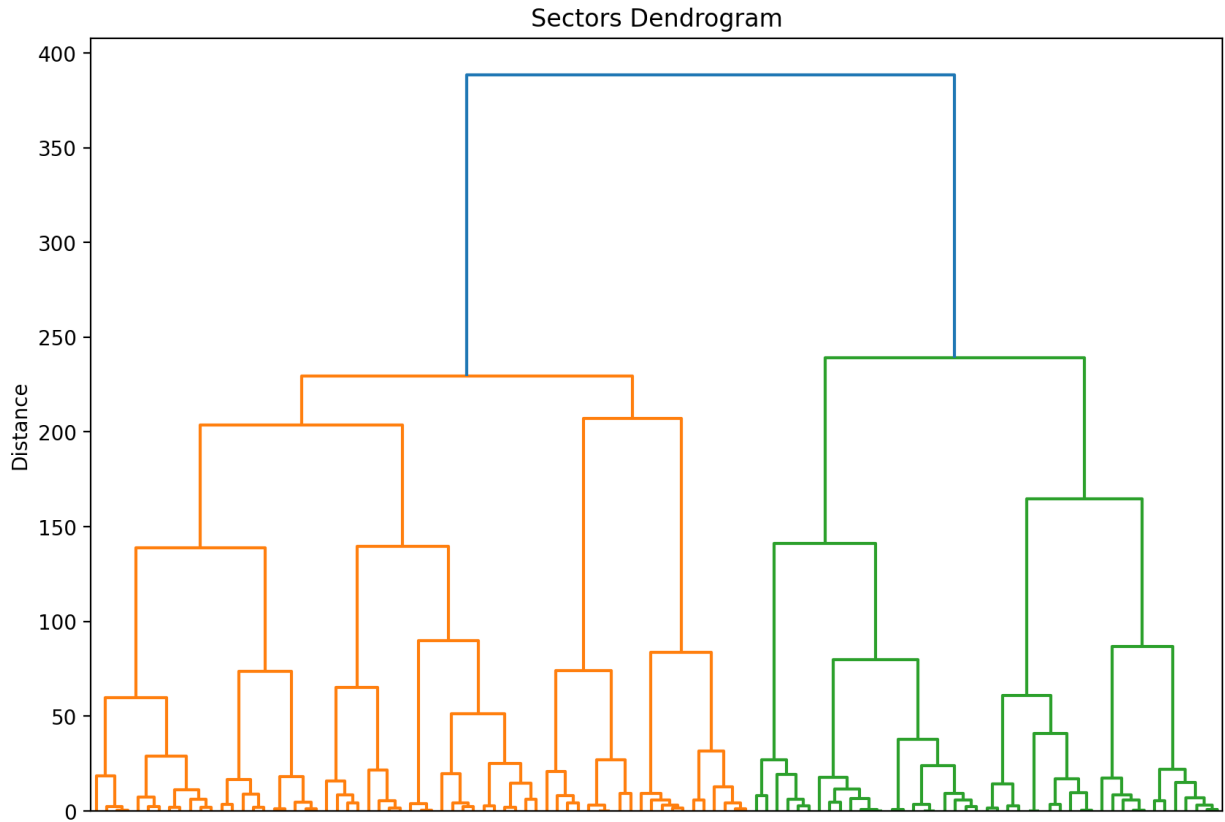

**Figure 4.** Dendrogram created from the DTW hierarchical clustering of the EF coefficients with the Sectors universe on the training data from 1999 to 2007. The dendrogram shows that there are two primary clusters in the Sectors universe.

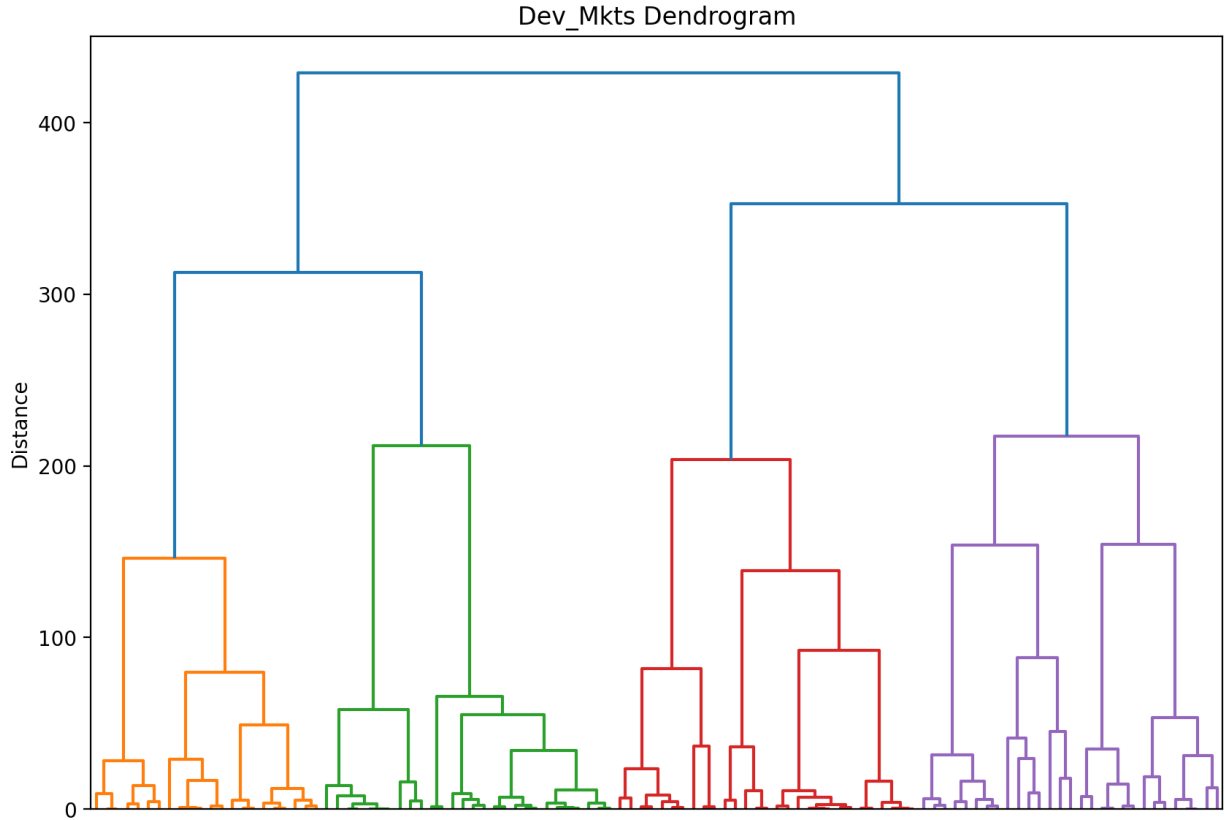

**Figure 5.** Dendrogram created from the DTW hierarchical clustering of the EF coefficients with the Developed Markets universe on the initial training data from 1988 to 1999. The dendrogram shows that there are four primary clusters for the Developed Markets universe.

In the markets, the hierarchical structure of regimes consists of a bull and bear state at the top level, and there are specific types of bull and bear states. The dendrograms all show that the data generally splits at either two or four clusters. Using only two states would not provide enough information, as we could simply use the sign of the average return to determine a two-state system. However, using too large a number of states would overfit, so we elected to use four clusters. To measure the sensitivity of the hyperparameter, we also ran the proposed model for three and five clusters. We ran the proposed model with three, four, and five clusters for each three universes, and for each run, held the number of clusters fixed over time to reduce the chances of over-tuning.

## 4. Backtest Methodology

Using the proposed model, we developed portfolios using three asset universes, and backtested in an online expanding fashion.

### 4.1. Ensuring Realistic Portfolios

To construct more realistic portfolios, we limited leverage, and added 1% transaction fees. To ensure that the tangency portfolios did not transfer all of the risk toward leverage, we added a constraint limiting the leverage to 50%. We added 1% daily transaction fees as this model would be used only on large and liquid assets, so the bid/ask spread will be narrow. Although the model generates new weights each month, the model is rebalanced daily to ensure that the weights do not drift from the target too much.

**4.2. Checking for Spurious Results**

Although we tested the model out-of-sample, the first two universes with which we tested only comprise U.S. equities, although this model should be generalizable to any market. To ensure that our results were not spurious, we chose to look at not only the two universes that partition the U.S. stock market, but also a completely different asset class: global macro indices in developed markets. Testing on a very different asset class is another form of out-of-sample testing that helps ensure our model is not overfit.

**4.3 Comparing to Benchmarks**

We measured the performance of our proposed model against three benchmarks: The tangency portfolio run monthly, the equal-weighed portfolio, and the S&P 500. The proposed Markov Markowitz model has lower volatility, and higher Sharpe ratio than the benchmarks as the state tangency portfolio can have long-short weights, which has lower overall risk from being market-neutral than the benchmarks, which are all long-only. To put the proposed model in the same risk units as the benchmarks for comparison, we volatility targeted the equal-weights portfolio.

**5. Empirical Results and Discussion**

**5.1. Backtest Performance of Proposed Model**

We will primarily focus on the Markov Markowitz model with four clustered states, as that best aligns with the dendrograms of the train data. We also ran the model with three and five clusters to measure the sensitivity of selecting the number of clusters. Figures 6-8 show that our proposed model with three, four, and five clusters all outperform the benchmarks with high return relative to volatility for each of the three universes.

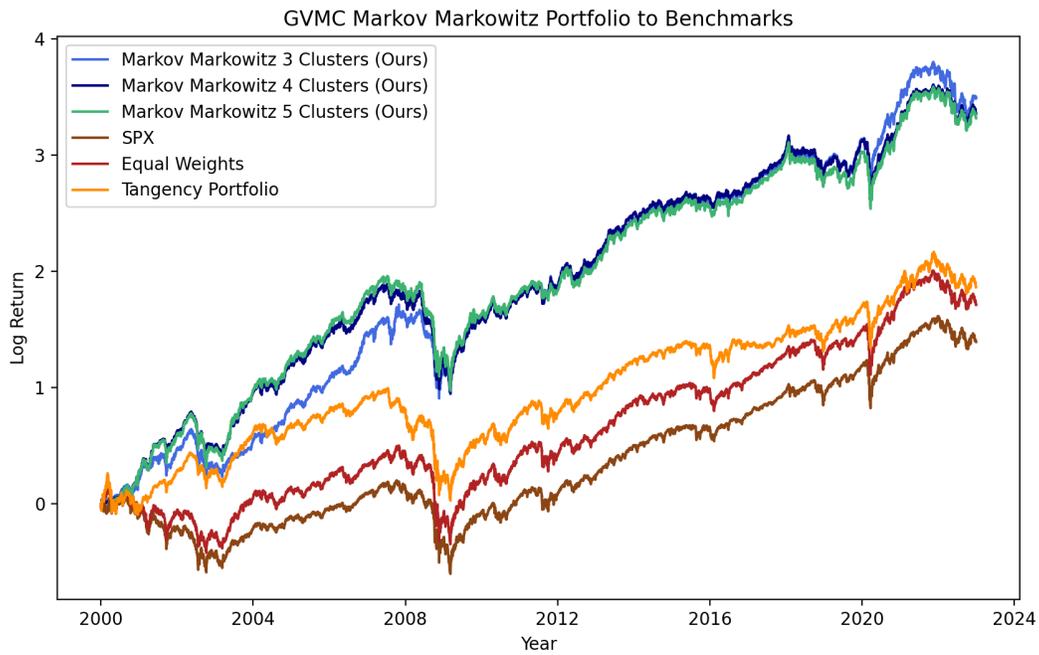

**Figure 6.** Performance of our proposed Markov Markowitz portfolio with 3-5 clusters as compared to benchmarks when run on the GVMC data out-of-sample from 2000-2022. The proposed models labeled "Ours" all outperform the benchmarks.

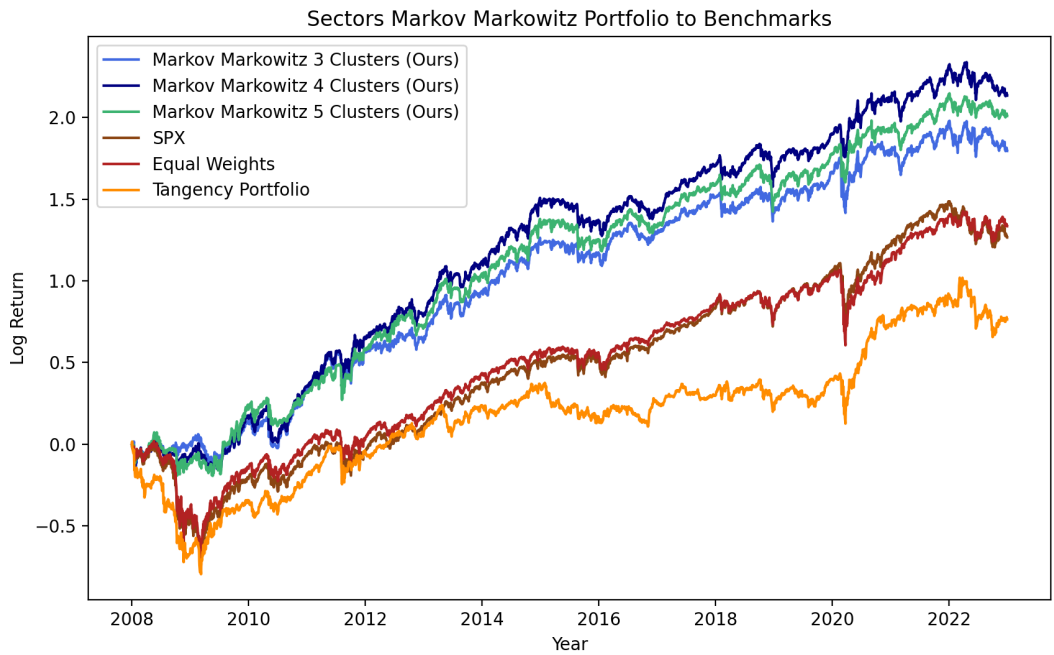

**Figure 7.** Performance of our proposed Markov Markowitz portfolio with 3-5 clusters as compared to benchmarks when run on the Sectors data out-of-sample from 2008-2022. The proposed models labeled "Ours" all outperform the benchmarks.

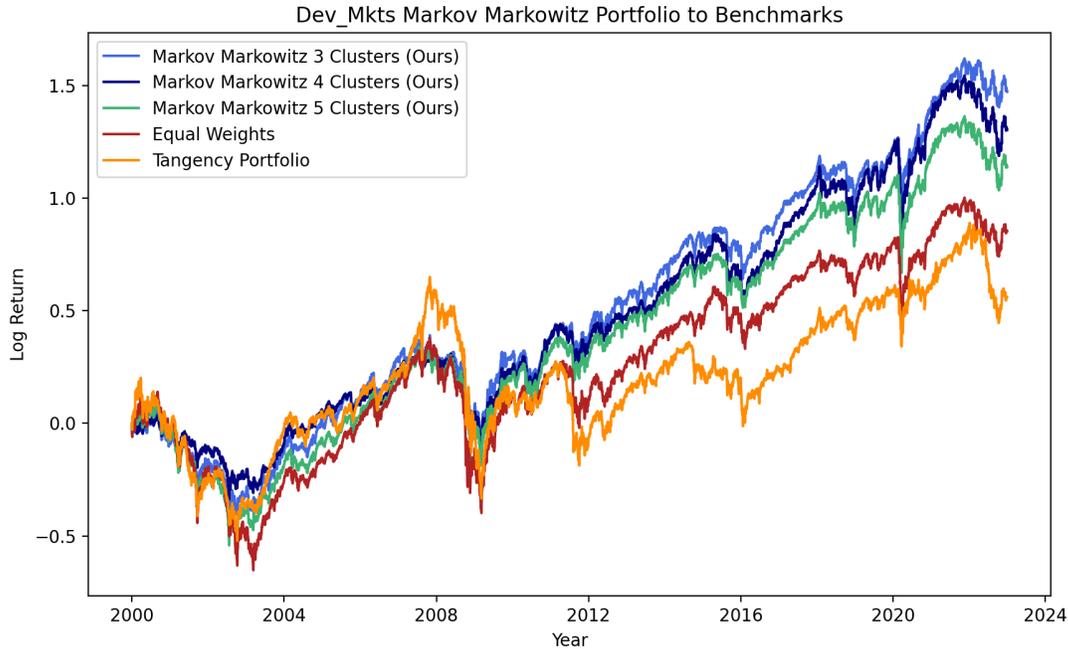

**Figure 8.** Performance of our proposed Markov Markowitz portfolio with 3-5 clusters as compared to benchmarks when run on the Developed Markets data out-of-sample from 2008-2022. The proposed models labeled "Ours" all outperform the benchmarks.

      For the GVMC and Developed Markets universes, using three clusters had the best performance, and for the Sectors universe, using four clusters had the best performance. We can see that the model is not highly sensitive to selecting the wrong number of clusters. We will show in section 5.3 that when the model is run with a higher number of clusters than necessary, the model will create multiple states with similar characteristics to each other, that would all be grouped together if the model was run with fewer clusters. However, unnecessarily splitting these clusters into a higher number of clusters does still reduces performance.

      We will focus on the performance of these models during the three shock market events during this period: the 2001 dot-com crash, 2008 Recession, and 2020 Covid crash. The proposed model performs especially well for the Sectors universe as it yields higher return while not suffering as large drawdowns from the three shock events as the benchmarks and the proposed model on the two other universes. This is likely because the Sectors universe has the more linearly independent assets than the other two universes, as the universe is sliced more finely with 9 assets, compared to 6 for GVMC and 5 for Developed Markets. For GVMC, although the model yields higher returns than the benchmarks, it still suffers from the same drawdowns for the three shock market events during this period. For the Developed Markets, the proposed model suffers lower

drawdowns during the Dot-com crash and 2008 Recession, but still suffers the same drawdown during the Covid crash.

The proposed model has a higher Sharpe ratio than the benchmarks and high annual return relative to max drawdown (MDD) as shown in Table 3 for each universe.

| Portfolio | GVMC (2000-2022) | | | Sectors (2008-2022) | | | Dev Mkts (2000-2022) | | |
|---|---|---|---|---|---|---|---|---|---|
| | Sharpe Ratio | Return Ann. (%) | MDD (%) | Sharpe Ratio | Return Ann. (%) | MDD (%) | Sharpe Ratio | Return Ann. (%) | MDD (%) |
| Markov Markowitz 3 Clusters | **0.66** | 19.0 | -55 | **0.63** | 15.1 | -29 | **0.46** | 7.9 | -41 |
| Markov Markowitz 4 Clusters | **0.64** | 18.3 | -61 | **0.73** | 17.7 | -23 | **0.42** | 7.1 | -40 |
| Markov Markowitz 5 Clusters | **0.63** | 18.1 | -63 | **0.69** | 16.7 | -25 | **0.38** | 6.4 | -48 |
| Tangency Portfolio | 0.43 | 10.6 | -62 | 0.33 | 7.4 | -56 | 0.22 | 4.2 | -63 |
| Equal Weighted | 0.39 | 10.1 | -57 | 0.51 | 11.6 | -51 | 0.30 | 5.1 | -55 |
| S&P 500 TR | 0.33 | 8.3 | -55 | 0.48 | 11.2 | -52 | - | - | - |

**Table 3.** Portfolio performance metrics of the proposed Markov Markowitz model for 3-5 clusters relative to benchmarks for each of the three universes. The proposed model with 3-5 clusters all have a higher Sharpe ratio than all benchmarks and high annual return relative to max drawdown for all three universes.

To establish that the outperformance is statistically significant, we performed alpha regressions on the proposed model with four clusters to each of the benchmarks. Table 2 show that the p-values for the alpha regressions with Jensen's alpha are all less than 0.05 for the GVMC and Sectors asset sets, so the outperformance of the proposed model is statistically significant to each benchmark. The developed markets assets have p-value less than 0.2, so the alpha is not statistically significant, but this is likely due to the total return not being included in the data, which limits the outperformance of the proposed model.

| Portfolio | GVMC | | Sectors | | Dev Markets | |
|---|---|---|---|---|---|---|
| | Annualized Alpha | P-value | Annualized Alpha | P-value | Annualized Alpha | P-value |
| Tangency Portfolio | 1.8% | 0.05 | 4.4% | 0.004 | 2.0% | 0.13 |
| Equal Weighted | 2.9% | 0.0001 | 3.1% | 0.02 | 1.0% | 0.20 |
| S&P 500 TR | 3.4% | 0.001 | 3.3% | 0.02 | - | - |

**Table 4.** Alpha regressions of the proposed Markov Markowitz portfolio with four clusters to benchmarks for each of the three universes.

## 5.2. Backtest Performance of Variations of the Proposed Model

Because the backtest is online expanding, the selection of the split should have minimal impact on performance. However, to ensure that the selection of the split did have any bias on the results, we also backtested the GVMC and Developed Markets universe initially split on 2008, which is the same split used for the Sectors universe. Figures 9 and 10 show that the proposed model still outperformed the benchmarks with the 2008 split. However, for GVMC, the proposed model with the 2008 split does not outperform the benchmarks to the same extent as with the 2000 split. Table 5 shows that proposed model with the 2008 split still had a higher Sharpe ratio than the benchmarks.

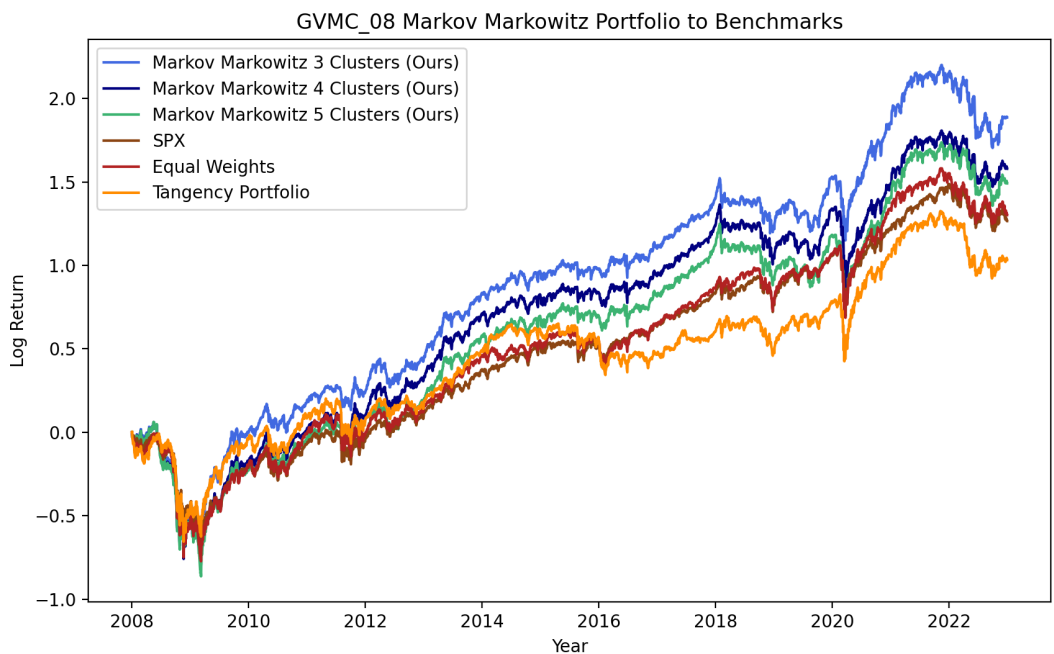

**Figure 9.** Performance of our proposed Markov Markowitz portfolio with 3-5 clusters as compared to benchmarks when run on the GVMC data out-of-sample from 2008-2022. The proposed models labeled "Ours" all outperform the benchmarks.

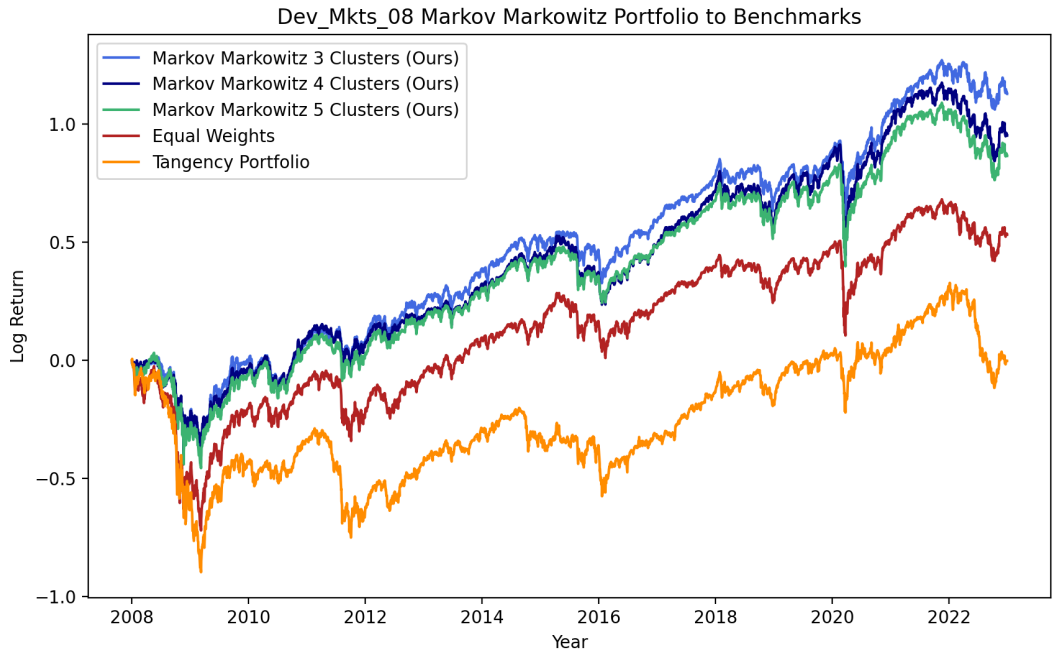

**Figure 10.** Performance of our proposed Markov Markowitz portfolio with 3-5 clusters as compared to benchmarks when run on the Developed Markets data out-of-sample from 2008-2022. The proposed models labeled "Ours" all outperform the benchmarks.

|  | GVMC (2008-2022) | | | Sectors (2008-2022) | | | Dev Mkts (2008-2022) | | |
|---|---|---|---|---|---|---|---|---|---|
| Portfolio | Sharpe Ratio | Return Ann. | MDD (%) | Sharpe Ratio | Return Ann. | MDD (%) | Sharpe Ratio | Return Ann. | MDD (%) |
| Markov Markowitz 3 Clusters | **0.62** | 16.1 | **-54** | **0.63** | 15.1 | -29 | **0.46** | 9.2 | -35 |
| Markov Markowitz 4 Clusters | **0.57** | 13.8 | -57 | **0.73** | 17.7 | -23 | **0.41** | 8.0 | -37 |
| Markov Markowitz 5 Clusters | **0.51** | 13.1 | -60.3 | **0.69** | 16.7 | -25 | **0.36** | 7.4 | -39 |
| Tangency Portfolio | 0.40 | 9.7 | -48 | 0.33 | 7.4 | -56 | 0.07 | 0.2 | -59 |
| Equal Weighted | 0.48 | 11.7 | -54 | 0.51 | 11.6 | -51 | 0.25 | 5.1 | -51 |
| S&P 500 TR | 0.48 | 11.2 | -52 | 0.48 | 11.2 | -52 | - | - | - |

**Table 5.** Portfolio metrics of the proposed Markov Markowitz model with 3-5 clusters and benchmarks for each of the three universes with the split starting in 2008. The proposed model with 3-5 clusters all have a higher Sharpe ratio than all benchmarks and high annual return relative to max drawdown for all three universes.

We also combined the assets in all of the universes and ran the proposed model with a 2008 initial split. Figure 11 shows that combining the universes yielded worse performance than each run individually. The higher dimensionality of the combined universe affects the estimation of the covariance matrices used in the proposed model. There are two different covariance matrices calculated in the Markov Markowitz model: the covariance matrices used in the mean-variance estimation and the covariance matrices used to calculate the monthly EF coefficients. The covariance matrices used in the mean-variance estimation has no estimation issues, as its over 10-year lookback provides significantly more data points relative to the 21 assets. However, the covariance matrix used to calculate the monthly EF coefficients only have about 21 business days of return data in each month, which is on average the same as the 21-asset dimensionality. When the number of assets is large relative to the number of return data points, as in this case, the sample covariance matrix tends to take on extreme values and is estimated with significant error (Ledoit and Wolf 2003). This estimation error likely contributes to the poor performance combined universe model. This suggests that the proposed model should only be used when the number of assets is fewer than the number of return data points in each time interval. To include a large number of assets, the covariance matrix used to calculate the EF coefficients would need to be calculated on intraday data, so that there are enough data points in each month.

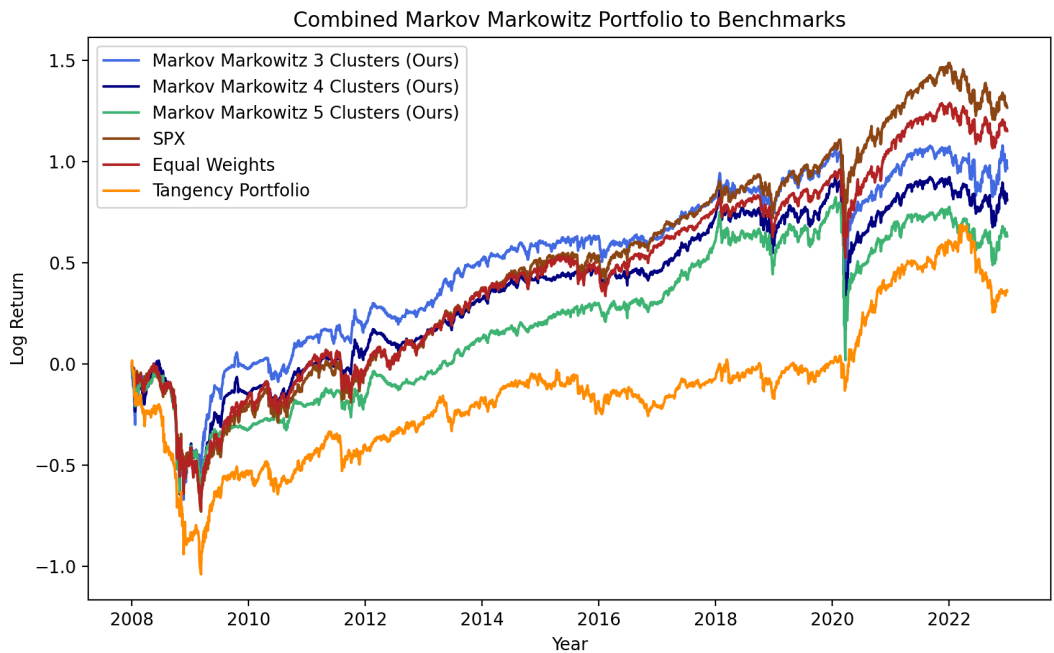

**Figure 11.** Performance of the Markov Markowitz portfolio with all three universes combined from 2008-2022. Although the proposed model outperformed all benchmarks for each universe separately, it underperforms the benchmarks when all three universes are combined. This is likely due to the lack of the degrees of freedom in the calculation of the monthly covariance matrix used to calculate the monthly EF coefficients.

### 5.3. Model Interpretability and Market Insights

A significant limitation with HMMs is the lack of interpretability due to the states being hidden. Because our proposed model uses observable states, we can observe intermediate components of the model to better understand how it determines the final weights. These intermediate components can also help ensure that the model follows intuition, as well as provide insight into the market. We will focus our analysis on the proposed model with four clusters, but a similar analysis could be performed for any number of clusters.

      We can visualize the clusters to see how well-separated they are. Figures 12-14 show the hierarchical clusters when trained on the entire period for the GVMC, S&P sectors, and Developed Markets assets. These clusters use all of the data, and the clusters created from online expanding framework would approach this final clustering of the entire dataset as the model iteratively updates with new data. Across all the dimensions, we can see that the clusters are well-separated. Unlike an HMM, we can clearly determine how the clusters were formed, as the clustering objective function only uses cluster distances, rather than attempting to fit clusters to a Markov model. This separation of the clustering and the Markov model also helps reduce the sensitivity of the proposed model to misestimation as compared to HMMs.

# GVMC Clustering of Efficient Frontier Coefficients

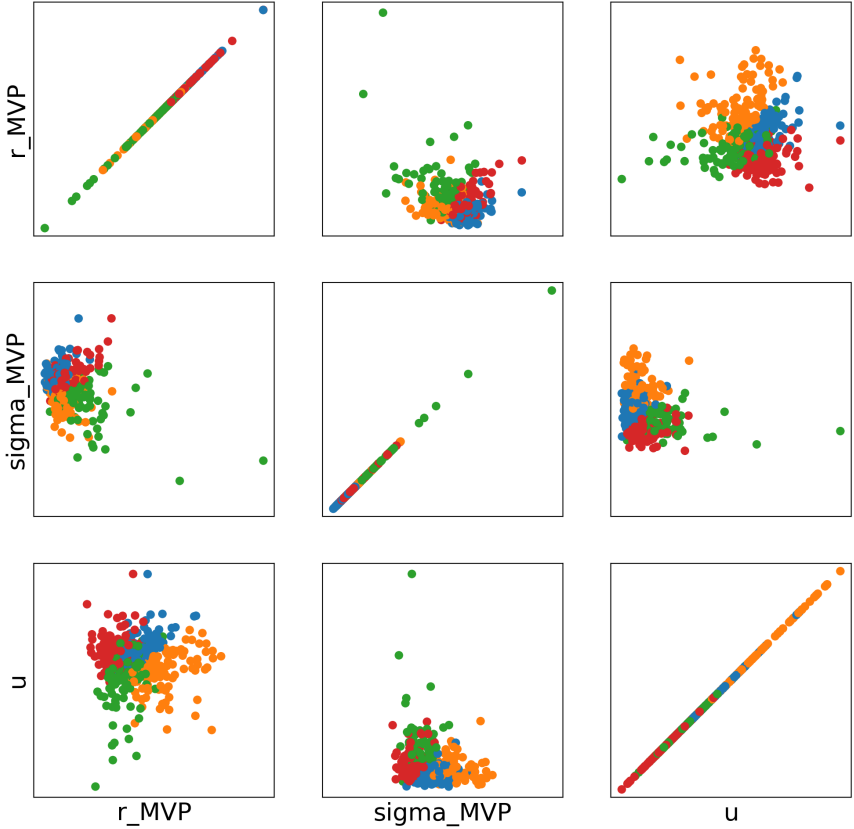

**Figure 12.** Scatterplots of the four DTW hierarchical clusters when run on the full GVMC data from 1990-2022. The clustering is performed across the three dimensions of each of the EF coefficients.

## Sectors Clustering of Efficient Frontier Coefficients

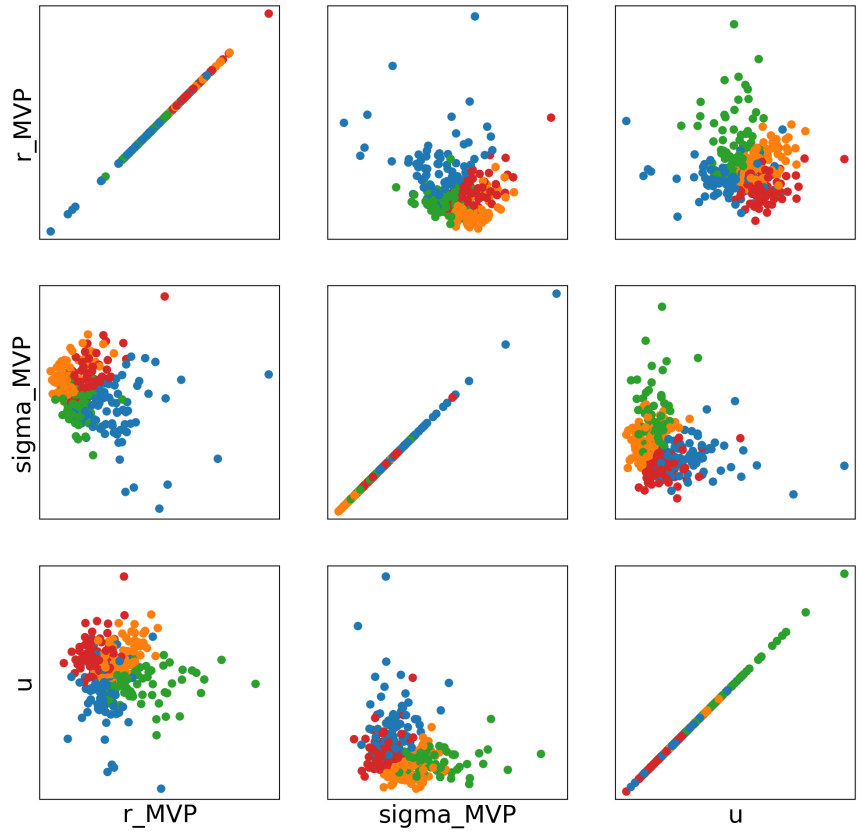

**Figure 13.** Scatterplots of the four DTW hierarchical clusters when trained on the full Sectors data from 1999-2022. The clustering is performed across the three dimensions of each of the EF coefficients.

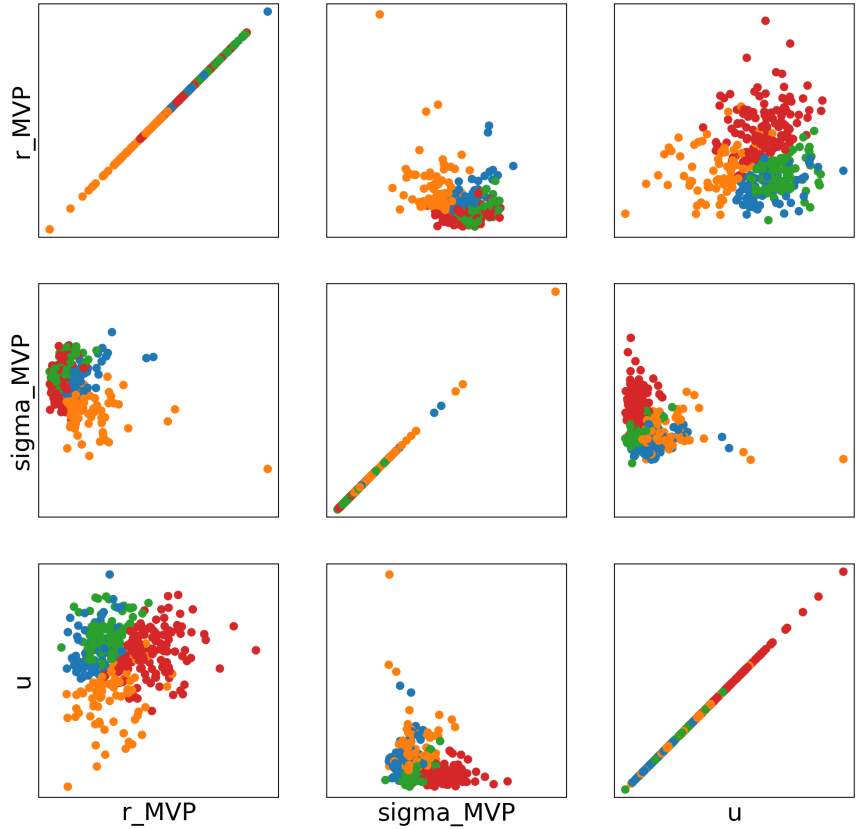

**Figure 14.** Scatterplots of the four DTW hierarchical clusters when trained on the full Developed Markets data from 1988-2022. The clustering is performed across the three dimensions of each of the EF coefficients.

Figures 15-17 provide visuals of the states over time for each set of assets and clustering methods. We color-coded the states based on the monthly recession indicators from FRED (2023). In the GVMC clusters (figure 15), state 3 includes most of the recessions, including the 1990 recession, the Dot-com bubble, 2008 financial crisis, and the Covid-19 market crash. Similarly, the Sectors clusters (figure 16) for state 1, and the developed markets clusters (figure 17) for state 2 include many of these crashes and recessions as well. When there is a recession, the broad market does not necessarily have low return for each month, as it could be a month where the market is recovering, which explains why the recessions are not grouped into one cluster.

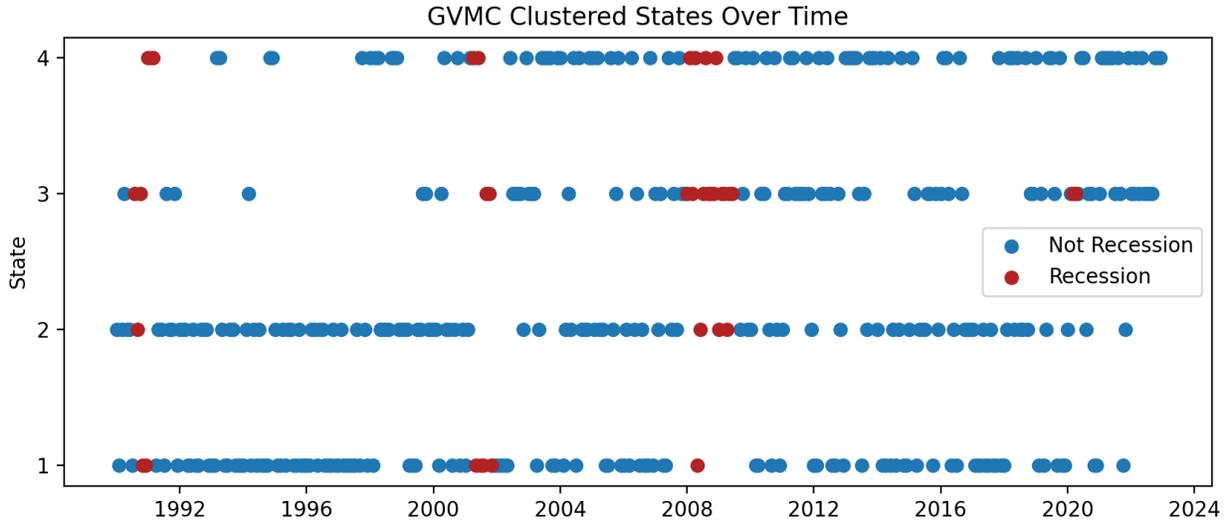

**Figure 15.** DTW hierarchically clustered states over time when trained on the full GVMC data from 1990-2022 color-coded by FRED recession indicators.

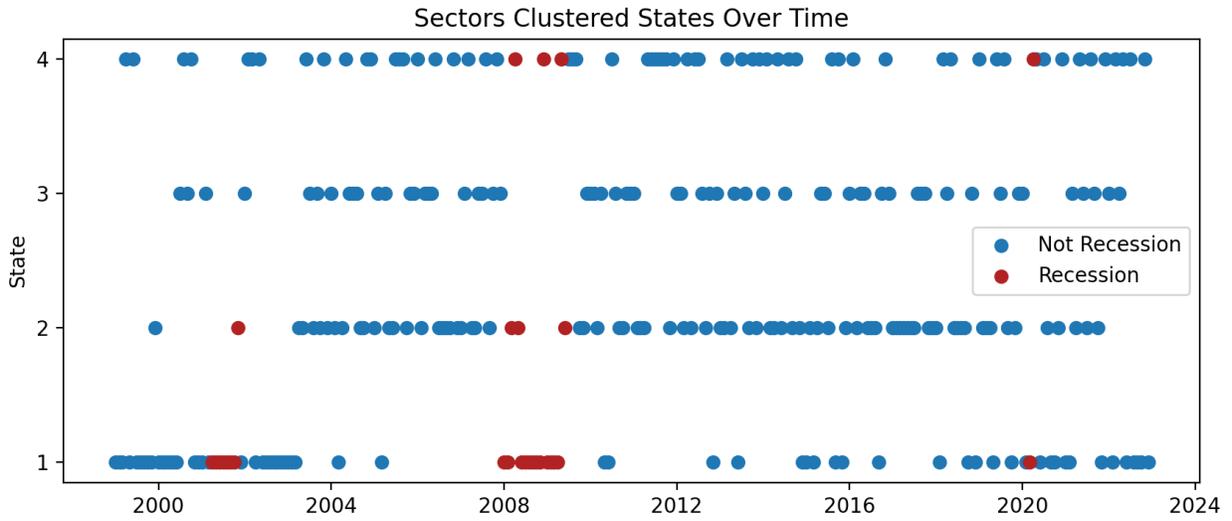

**Figure 16.** DTW hierarchically clustered states over time with the trained on the full Sectors data from 1999-2022 color-coded by FRED recession indicators.

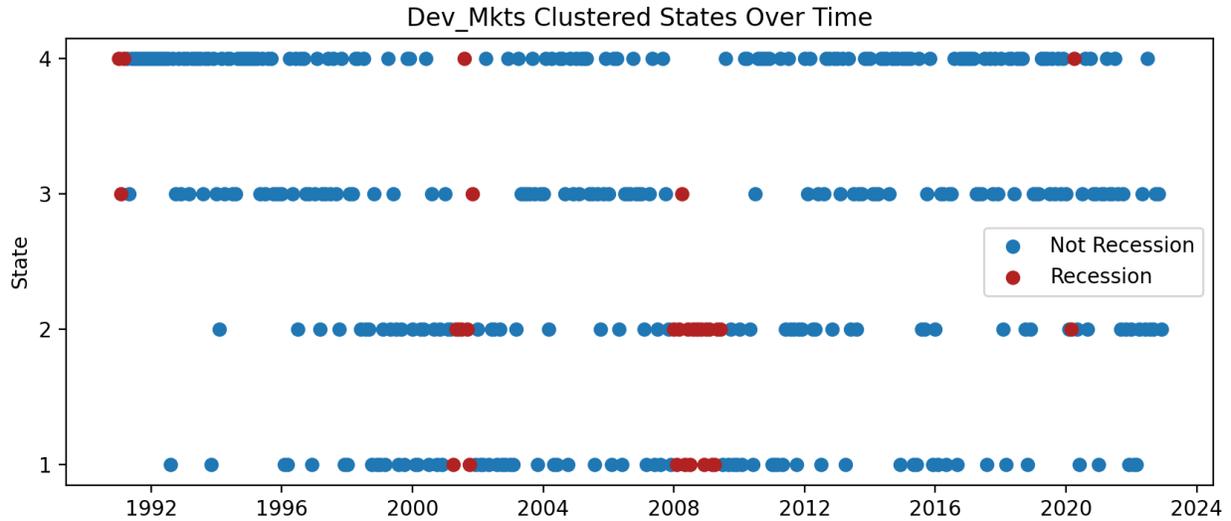

**Figure 17.** DTW hierarchically clustered states over time when trained on the full Developed Markets data from 1988-2022 color-coded by FRED recession indicators.

We can compare these clusters to the market to understand the extent to which the clustering was driven by market return and volatility. Figures 18-20 provide the average equal-weighted market return and standard deviations within each of the states. State 3 in the GVMC clusters (figure 18), state 1 in the sectors clusters (figure 19), and state 2 in the developed markets clusters (figure 20) have negative broad market return, and high volatility, which aligns with how these states includes many market crashes and recessions. We can also see for the GVMC (figure 18) and Sectors (figure 19), there is one bearish state, two bullish states, and one state with low return. However, for the Developed Markets (figure 20), there is one bearish state, and three bullish states. Among the three bullish states, one has high market return, and two have moderate market return.

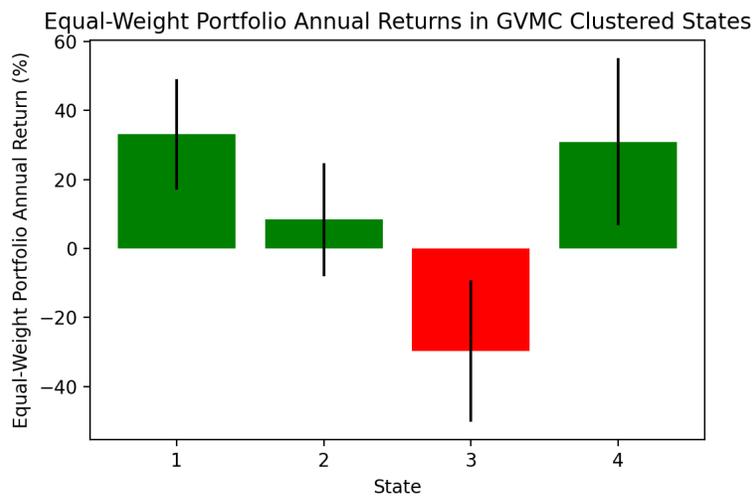

**Figure 18.** The equal-weight portfolio average annualized return and standard deviation within each of the clustered states trained on the full GVMC data from 1990-2022. For example, the green bar for state 1 represents the average return across all months labeled as state 1 for all assets.

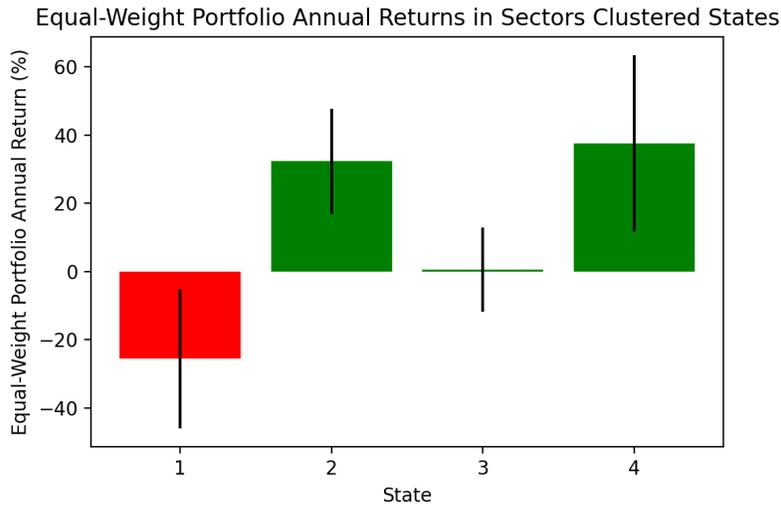

**Figure 19.** The equal-weight portfolio average annualized return and standard deviation within each of the clustered states when trained on the full Sectors data from 1999-2022. For example, the red bar for state 1 represents the average return across all months labeled as state 1 for all assets.

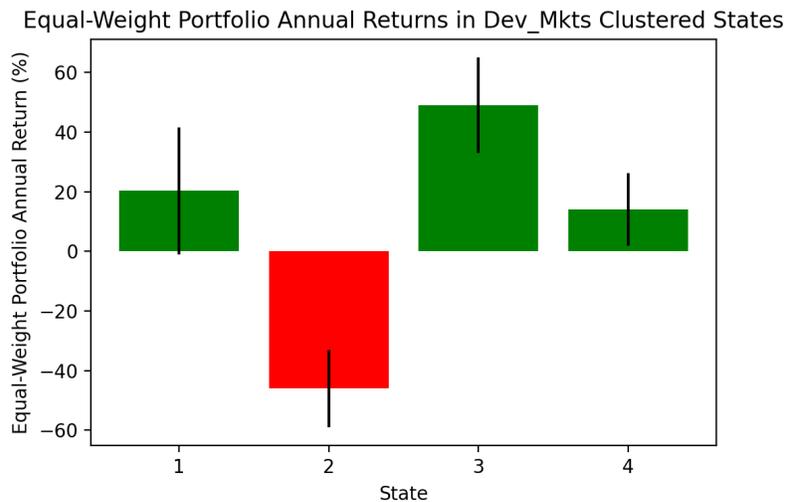

**Figure 20.** The equal-weight portfolio average annualized return and standard deviation within each of the clustered state when trained on the full Sectors data from 1988-2022. For example, the green bar for state 1 represents the average return across all months labeled as state 1 for all assets.

After observing the clustered states, we can observe the matrix of weights for each state. The heatmaps of the weights in each state in Figures 21-23 show that portfolios in each state are significantly different from each other, as they are tailored to the risk and reward opportunity in that state.

Figure 21 shows that the GVMC universe has two net long states and two market-neutral states. In the bearish cluster, state 3, the portfolio is market-neutral, going long mid-cap and short large growth and small-cap value. State 2, which has low market return, is similarly market-neutral, but instead goes long large-cap growth and short large-cap value. States 1 and 4, the two more bullish states, are net long and each put the majority of weight in one asset, mid-cap value, but hedge slightly differently. Overall, states 1 and 4 are very similar, suggesting the true model for GVMC universe would only have three states, where states 1 and 4 are combined.

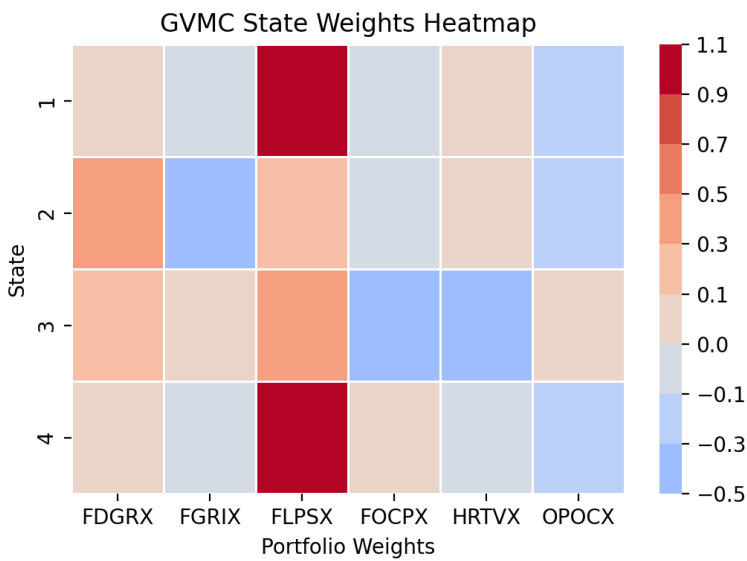

**Figure 21.** Heatmap of the Markov Markowitz weights in each of the clustered states when trained on the full GVMC data from 1990-2022. Each row corresponds to the weights of the tangency portfolio calculated using only return data clustered into that row's state. The weights can either add to 1 or 0, whichever maximizes the Sharpe ratio, and always have no more than 50% leverage.

Figure 22 shows that the Sectors universe has two net long states and two market-neutral states. In the bearish state, state 1, the portfolio is heavily short the materials and industrial sectors, while long consumer goods and health-care, and is market-neutral. State 3, which has low market return, is market-neutral, while states 2 and 4, the bullish states, are net long. The allocations of the two bullish states are significantly different. State 2 allocates close to equal weight to all sectors except Materials and Energy, which have negligible weight, and the Finance sector is shorted. State 4 is instead more selective, similarly being long Industrials, Utilities and Healthcare, and short the Finance, but instead also being long materials, and having negligible weight on the remaining sectors. As there are no two sets of weights that are particularly similar, this suggests that the true model has at least four states for the Sectors data.

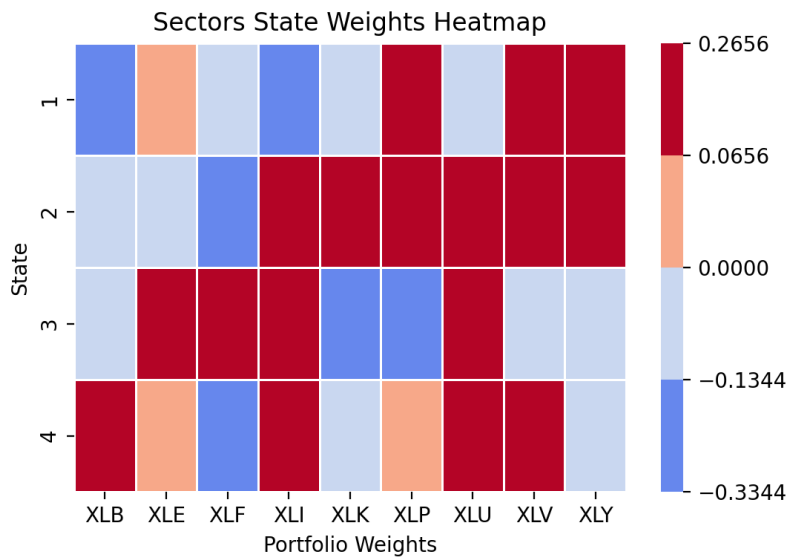

**Figure 22.** Heatmap of the Markov Markowitz weights in each of the clustered states over time trained on the full Sectors data from 1999-2022. Each row corresponds to the weights of the tangency portfolio calculated using only return data clustered into that row's state. The weights can either add to 1 or 0, whichever maximizes the Sharpe ratio, and always have no more than 50% leverage.

Figure 23 shows that the Developed Markets universe has two net long states (1 and 3) and two market-neutral states (2 and 4). In state 2, the bearish state, the portfolio heavily shorts the German DAX, while going long the Canadian TSX. The heavily bullish state, state 3, allocates positive weights across all assets. States 1 and 4, the two bullish states with moderate return, takes the largest long position in the S&P 500, but hedge slightly differently. Overall, states 1 and 4 are very similar, suggesting the true model for Developed Markets universe would only have three states, where states 1 and 4 are combined.

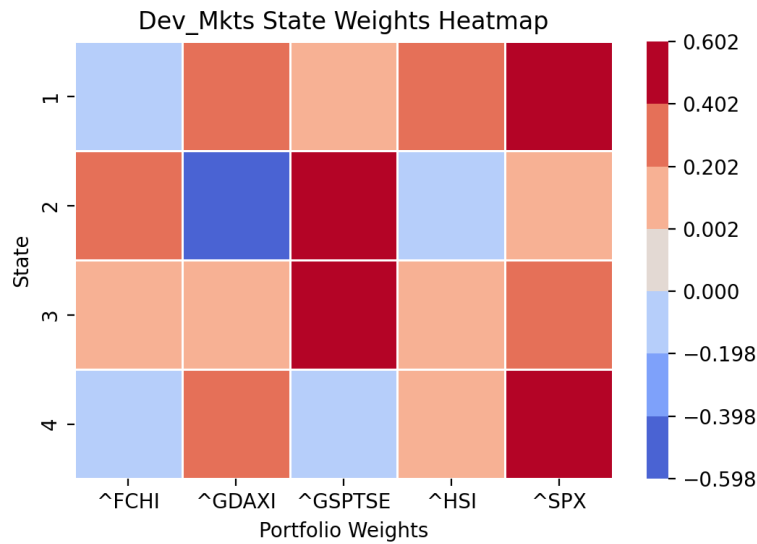

**Figure 23.** Heatmap of the Markov Markowitz weights in each of the clustered states over time when trained on the full Developed Markets data from 1988-2022. Each row corresponds to the weights of the tangency portfolio calculated using only return data clustered into that row's state. The weights can either add to 1 or 0, whichever maximizes the Sharpe ratio, and always have no more than 50% leverage.

Finally, the Markov state-transition matrices in tables 6-8 provide insight into how the market states move. Overall, for all of the universes, one takeaway is that one of the bullish states is always significantly more likely to lead to a bearish state than the other non-bearish states.

Table 6 shows that in the GVMC universe, all states have a 30-40% probability of recurring. In addition, state 4, one of the bullish states, is almost two times more likely to lead to the bearish state, state 3, than the other two non-bearish states at 22%. The bearish state, state 3, also has a higher probability or recurring than transitioning to any other state at 37%.

|  | Transition State | | | |
| --- | --- | --- | --- | --- |
| **Initial State** | **1** | **2** | **3** | **4** |
| **1** | 0.41 | 0.30 | 0.13 | 0.15 |
| **2** | 0.36 | 0.30 | 0.13 | 0.21 |
| **3** | 0.14 | 0.25 | 0.37 | 0.24 |
| **4** | 0.23 | 0.21 | 0.22 | 0.34 |

**Table 6.** State-transition matrix of the DTW hierarchically clustered market states with the full GVMC data from 1990-2022.

Table 7 shows that in the Sectors universe, the bearish state, state 1, is almost twice as likely to recur as compared to the other states. In addition, the most likely state to lead to the bearish state is state 4, one of the bullish states, which is almost twice as likely than the other two non-bearish states at 26%. The bearish state, state 1, also has a significantly higher probability or recurring than transitioning to any other state at 56%.

|  | Transition State | | | |
| --- | --- | --- | --- | --- |
| **Initial State** | **1** | **2** | **3** | **4** |
| **1** | 0.56 | 0.16 | 0.11 | 0.16 |
| **2** | 0.12 | 0.35 | 0.23 | 0.29 |
| **3** | 0.14 | 0.32 | 0.27 | 0.27 |
| **4** | 0.26 | 0.32 | 0.23 | 0.20 |

**Table 7.** State-transition matrix of the DTW hierarchically clustered market states with the full Sectors data from 1999-2022.

Table 8 shows that in the Developed Markets clustering, unlike the other two asset universes, the bearish state, state 2, does not have a higher probability of recurring than transitioning to another state. In addition, state 1, one of the bullish states, has more than 3 times the probability of transitioning to the bearish state, state 2, at 42%.

| Initial State | Transition State | | | |
|---|---|---|---|---|
| | 1 | 2 | 3 | 4 |
| 1 | 0.19 | 0.42 | 0.16 | 0.23 |
| 2 | 0.32 | 0.23 | 0.19 | 0.27 |
| 3 | 0.16 | 0.12 | 0.30 | 0.41 |
| 4 | 0.15 | 0.10 | 0.31 | 0.45 |

**Table 8.** State-transition matrix of the DTW hierarchically clustered market states with the full Developed Markets data from 1988-2022.

The steady-state probabilities in table 6 provide the long-run probabilities of ending up in each state. For all three universes, there is a less than 1/3 probability of ending up in the bearish state. The lowest steady-state probability is 0.19, so there are no states that are particularly unlikely to not end up in.

| Assets Set | State | | | |
|---|---|---|---|---|
| | 1 | 2 | 3 | 4 |
| GVMC | 0.30 | 0.27 | 0.20 | 0.23 |
| Sectors | 0.28 | 0.29 | 0.21 | 0.23 |
| Dev Mkts | 0.19 | 0.19 | 0.25 | 0.36 |

**Table 9.** Steady-state probabilities of the DTW hierarchically clustered states from each of the three universes. The probabilities highlighted in orange are the probabilities of ending up in the bearish state.

## 6. Conclusions and Future Research

This paper presents a novel asset allocation approach using a Markov process of clustered states described by EF coefficients. Our approach uses a decomposition of EFs into three interpretable coefficients: $r_{MVP}$, $\sigma_{MVP}$, and $u$. The proposed method uses hierarchical clustering to cluster states defined by these monthly EF coefficients. This approach then maps each state to a tangency portfolio calculated using data only in that state, unlike standard tangency portfolios, which would use all the data regardless of state with some lookback. Then, the model takes the expectation of these weights by the probability of transitioning to the associated state based on the Markov model. Using this method out-of-sample from 2000-2022 and 2008-2022 for three asset universes, this model outperformed three benchmark portfolios.

This method is interpretable, as it allows users to observe the underlying clustered states and Markov process structure. Upon inspecting the intermediate components of the model, we found a number of insights into the market. Most notably, we found that among each of the universes, there exists multiple bull market states, with one significantly more likely to transition to a bear market than the others. In addition, we found that for the universes that partition the U.S. stock market, when in the bearish state, it is more likely to recur than transition to any other state. However, this recurrence property does not hold for the Developed Markets universe.

Although this analysis yielded significant results, there are still certain limitations that we will investigate in future research. As we did not have total return data for the developed markets, the outperformance was not as significant, so in the future we plan to acquire developed markets total returns data to test. We will also test the model on other sets of assets that span the market beyond the three sets we tested, such as commodities, FX, and fixed income, as our results could have been dependent on the selected assets. For future work, we will also define daily states using intraday data to test the proposed model on a higher frequency with a larger set of assets. Future research will also compare the clustered states to historical financial events beyond our brief observation of the market crash and bull market state clusters, to determine implications of the state transitions from these events.